\documentclass[10pt,twocolumn]{article}
\usepackage{graphicx,epsfig,url,float}
\begin{document}

\title{Robust seasonal cycle of Arctic sea ice area through tipping point in amplitude.}
\author{Peter D. Ditlevsen\\ Centre for Ice and Climate, Niels Bohr Institute\\Juliane Maries Vej 30, DK-2100 Copenhagen O, Denmark}

\maketitle

\twocolumn

{\bf
The variation in the Arctic sea ice is dominated by the seasonal cycle with little inter-annual correlation. Though 
the mean sea ice area has decreased steadily in the period of satellite observations,  a dramatic transition in the 
dynamics was initiated with the record low September ice area in 2007. The change is much more pronounced in the amplitude 
of the seasonal cycle than in the annual mean ice area. The shape of the seasonal cycle is surprisingly constant for the whole 
observational record despite the general decline. A simple
explanation, independent of the increased greenhouse warming, for the shape of the seasonal cycle is offered. Thus the
dramatic climate change in arctic ice area is seen in the amplitude of the cycle and to a lesser extend the annual mean and
the summer ice extend. The reason
why the climate change is most pronounced in the amplitude is related to the rapid reduction in perennial ice and thus a
thinning of the ice. The analysis shows that  a tipping point for the arctic ice area was crossed in 2007.
}

\section*{Introduction}
Observations[\cite{cavalieri:2012}] show more dramatic decline in the ice area as a response to the increased greenhouse gas forcing than predicted by climate model simulations[\cite{holland:2006}]. Most dramatic is the decrease in the September minimum  ice extend[\cite{stroeve:2012,cavalieri:2012}]. 
The projections of the Arctic sea ice are important for the perspective of future exploitation and possibility of commercial shipping through the Arctic Ocean, where
completely ice free summers have been suggested as soon as 2030[\cite{stroeve:2007}]. But more importantly, the reduction in the sea ice might, through the summertime ice
albedo feedback, play an essential role in the polar amplification of global warming[\cite{screen:2010,polyakov:2002,holland:2003,johannessen:2004}]. The sea ice extend depends on many factors, such as changed atmospheric circulation patterns[\cite{dickson:2000}], which together with changed heat fluxes in the open ocean, more clouds and water vapor[\cite{francis:2006}] raise the surface-air temperature (SAT)[\cite{screen:2010}] and thus increase summer melt[\cite{lindsay:2009}]. Other factors are increased wind driven ice export through the Fram Strait[\cite{ogi:2012}] or changes in location and strength of the Beaufort Gyre[\cite{proshutinsky:2009}]. The ice-albedo feedback in the melt
season is not well understood, it response non-linearly to the change in ice ages[\cite{perovich:2007,fowler:2004}].  Furthermore, the effect of the ice albedo also depends on the cloud formation over breakup areas[\cite{kay:2009}]. The poorly constrained dependency of the sea ice on all these factors[\cite{serreze:2007}], the large natural variability and the limitations in understanding the 
effect of climate change on patterns like the North Atlantic Oscillation (NAO) and the Northern Annular Mode (NAM)[\cite{dickson:2000,rigor:2002}] makes projections difficult.
Even though there is broad consensus that the positive feedback from the ice albedo contributes to warming in the Arctic, there is much less agreement on
it's role in comparison to other processes in the polar amplification of global warming[\cite{serreze:2006,winton:2006}].      

In order to better understand how the Arctic sea ice has reacted to climate change, an analysis of the observed variations in the total ice area is performed.  
The sea ice area is derived from NASAÕs Satellite based Scanning Multichannel Microwave Radiometer (SSM/I and SMMR)[\cite{cavalieri:1996}] and
obtained from the University of Illinois' project "The Cryosphere Today"[\cite{thecryospheretoday}]. The observed record (figure 2, top panel), covers the Arctic Ocean and
surrounding waters in the period 1979-present with daily resolution.

\section*{Decomposing the Annual Cycle}
The resent years, 2007 to present, show a remarkable decrease in annual mean (red and green curves in figure 1 (a)), with a
pronounced summer melting. In order to quantify the climate change and change to the seasonal cycle the
sea ice area can be described by $$x_i(t)=m_i+A_i f(t)+\epsilon_i(t),$$ where $i\in(1979,2011)$ denotes year, and $t\in (1, 365)$ denotes day of year.
\begin{figure}[H]
\begin{center}
\epsfxsize=8cm\epsffile{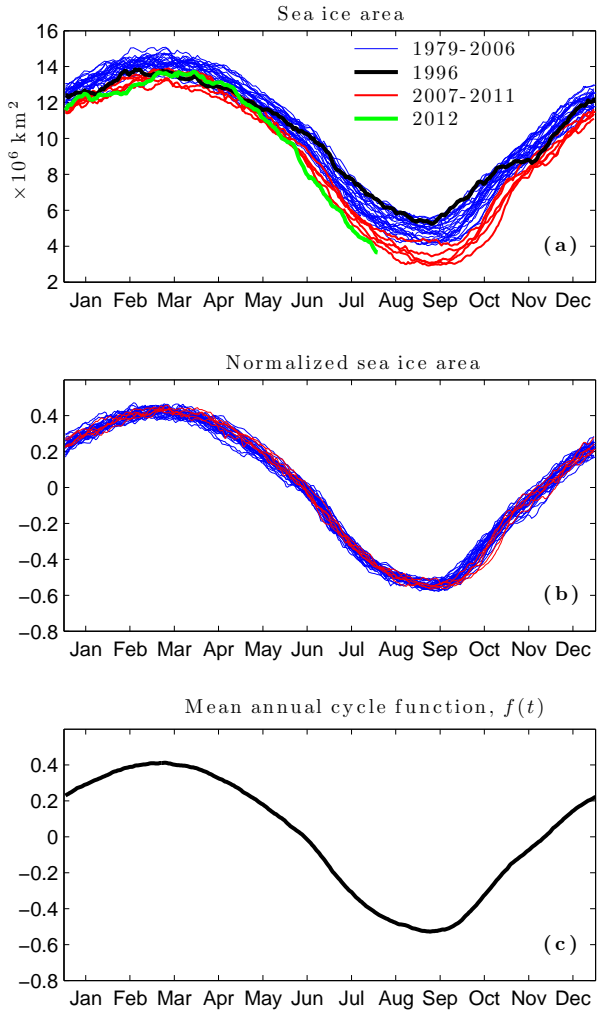}
\end{center}
\caption{The seasonal variation of the Arctic sea ice area. (a) shows all the years obtained from the satellite measurements. A year round decline in is seen after 2007. 
The year 1996 showed a notably small amplitude in the seasonal cycle. (b) Same data as in (a) but normalized by subtracting annual mean and dividing by the 
seasonal amplitude (2012 excluded). There is a striking collapse of all years despite the pronounced climate change after 2007. (c) The mean annual cycle function is obtained as
the day-by-day average of the 33 normalized curves in (b).  
}
\label{f1}
\end{figure} 
The function $f(t)$ has zero mean and represents the (constant) seasonal cycle. The mean $m_i$ and amplitude $A_i$ are constant within
a given year $i$. The residual $\epsilon_i(t)$ is assumed to be a simple stochastic noise. It is not given a priory that this is an adequate 
description of the Arctic sea ice area. However, by plotting the normalized ice area $(x_i(t)-m_i)/A_i=f(t)-\epsilon_i(t)/A_i$ we see an almost perfect collapse
of the data (figure 1(b)), see the appendix for how to estimate $m_i$ and $A_i$. Thus the function $f(t)$ can be accurately
estimated as the the mean of the normalized ice area for all years (figure 1(c)).  As the difference between the two lower panels
in figure 1 is small, the residual noise $\epsilon_i(t)$ is small. This is plotted with the ice area in figure 2(a) (2012 cannot be included,
since the data does not cover the full year, however, see supplementary material: www.gfy.ku.dk/$\sim$pditlev/seaice-suppl.pdf). Two findings are surprising regarding the residual noise: Firstly, despite the dramatic 
reduction in Arctic sea ice through the record, there are no trends in the noise. Secondly, the residual noise is a perfect red noise with
a correlation time of $\tau=22$ days, since the autocorrelation is exponential (figure 2(b)) and the compensated noise $\tilde{\epsilon}(t)=\epsilon(t)-\exp(-1/\tau)\epsilon(t+1)$ is 
perfectly uncorrelated and structureless (figure 2(c)). The correlation time of 22 days is consistent
with the time scale of variations in sea ice area governed by the natural atmospheric variability. The annual mean $m_i$ (figure 2(d)), shows,
as has been reported before[\cite{stroeve:2012,lindsay:2009,comiso:2008,maslanik:2007}], a downward trend through the full record, while the amplitude of the seasonal cycle $A_i$ (figure(2(d)) has a distinct
minimum for 1996 and a sudden positive jump in 2007, from which it has not recovered, indicating a new state dominated by one-year ice[\cite{maslanik:2007}]. (See the appendix for a discussion on the statistical significance).
\begin{figure}[H]
\begin{center}
\epsfxsize=8cm\epsffile{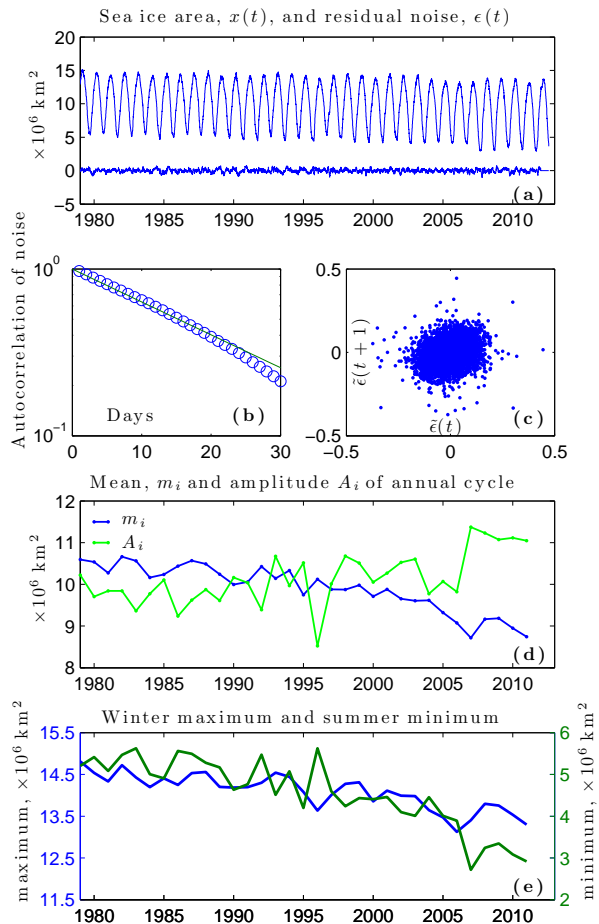}
\end{center}
\caption{The decomposition of the Arctic sea ice are. (a) shows the satellite measurements since 1979, with the small residual noise shown on the same scale. (b) The
autocorrelation of the residual noise is almost perfectly exponential. The green line is the curve $\exp(-t/\tau)$, with $\tau=22$ days. (c) The scatter plot of the compensated 
noise (see text for explanation) shows that there is no structure, beside the simple exponential autocorrelation in the residual noise. (d) shows the steadily declining mean 
ice area $m_i$ and the amplitude of annual cycle, which beside the outlier at 1996 shows an increase, with a jump in 2007. (e) shows the winter maximum and summer minimum (observe shifted axis for comparison), these are obtained from the mean and amplitude as $min_i=m_1+A_i min(f(t))$ and $max_i=m_i+A_i max(f(t))$ which
occur on September 8. and March 9., respectively. 
}
\label{f2}
\end{figure} 

The sea ice area at any particular day of the year is easily obtained from the
decomposition, thus we have the summer minimum $min_i=m_i-0.53 A_i$, where $min(f(t))=f(\mbox{September 8})=-0.53$ and the winter maximum $max_i=m_i+0.41A_i$,
$max(f(t))=f(\mbox{March 9})=0.41$. These are shown in figure 2(e), both with a downward trend, but not as clear a jump in 2007 as is seen in the amplitude $A_i$.
Note that the natural variability unrelated to climate change, represented by $\epsilon_i(t)$ is filtered out of the estimates for $min_i$ and $max_i$.  

\section*{The Tipping Point}
A change in the linear trend in $m_i$ as well as in summer minimum can be inferred from
linear regression analysis, though with low statistical significance[\cite{comiso:2008}].
Here we shall implement a Bayesian inference change point analysis, which is a slightly more advanced method
for detecting changes in statistics. This analysis, see the appendix for details, tests the data for probable changes in statistics. It is assumed
that the data follows gaussian distributions with constant mean and variance between (unknown) change points. The likelihood of the
realized data after the previous change point is calculated and compared to the likelihood if a change of mean and variance has occurred at a new change point.
This probability is shown in grey scale in figure 3(c), where white corresponds to probability 0 and black to probability 1. The red curve tracks the maximum probability.
The contour plot is read in the following way: A change point is defined at the beginning of the record, thus for a given year, say 1994, the maximum
probability of the duration since the last change point is 15 years (time since 1979). Probability of duration longer than 15 years is obviously zero (lower white
triangle). A change point in the statistics for the amplitude $A_i$ is detected at 2007. The cumulated distribution function for the amplitude prior to the change point
is shown in figure 3(d), it follows closely a gaussian distribution (black curve). Note that the 1996 low amplitude is a $3\sigma$ event, thus in figure 3(c) a slightly higher
probability of a changing point at 1996 is seen. The corresponding analysis for the mean $m_i$ (not shown) does not detect any changing points in
the statistics. Thus the amplitude of the seasonal cycle $A_i$ is the better fingerprint of Arctic climate change. A scatter plot of $A_i$ vs. $m_i$ (figure 3(e)) clearly
shows a change after 2007, which has not recovered (color coding is the same as in figure 1(a). Note again the 1996 outlier (black)). With the detection of the changing point,
the mean $m_i$ is shown in figure 3(a) with tendency for the two periods obtained from linear regression (blue lines), in figure 3(b) the amplitude is shown with the 
mean values for the two periods (blue lines). In both cases the green dashed line shows the tendency (for $m_i$) or mean values (for $A_i$) when no change in statistics is assumed. 
\begin{figure}[H]
\begin{center}
\epsfxsize=8cm\epsffile{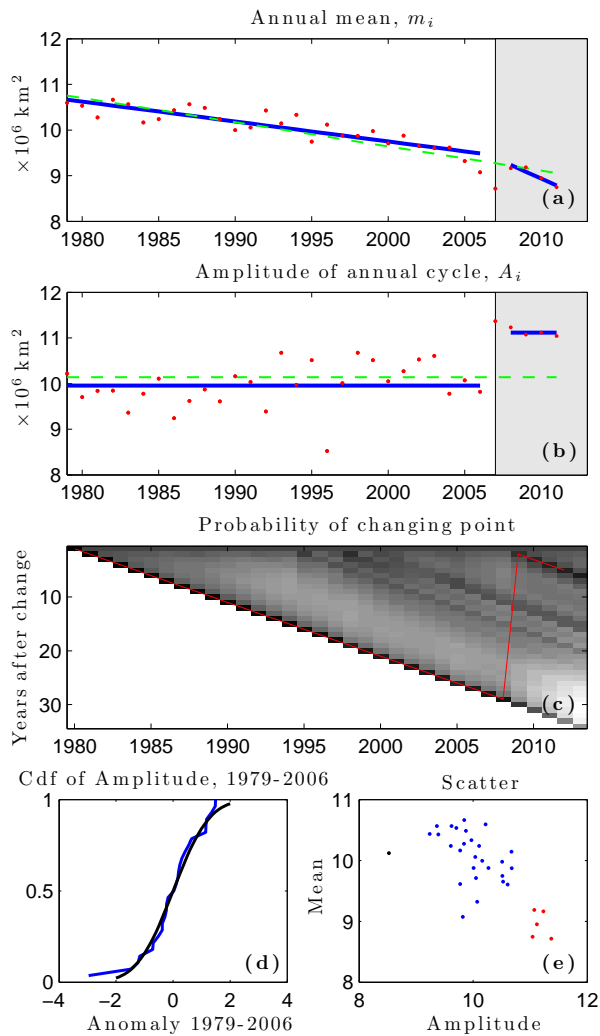}
\end{center}
\caption{
The change in statistics at 2007. (a) shows an increase in the linear trend after 2007, which showed a deep minimum. Beside the general trend (green
dashed curve) the change in the slope after 2007 (blue lines) is statistically marginal. (b) The change in the amplitude in 2007 is much more significant. The last
five points are much higher than the mean for the period (green dashed line) indicating a change in mean (blue lines). (c) The result of a Bayesian inference of
change in statistics of the amplitude of the annual cycle. A change point is detected in 2007 (see text for explanation). (d) shows that the annual amplitude prior to
2007, normalized to unit variance followed a gaussian distribution (black curve). Note that 1996 is a $3\sigma$ outlier. (e) Scatter plot of $m_i$ vs. $A_i$ where
the color coding is the same as in figure 1(a). This indicates that 2007 is a tipping point to a new statistical state. Note that $A_i$ and $m_i$ are independent 
within the two populations, prior to 2007 (blue and black points) and after 2007 (red points).   
}
\label{f3}
\end{figure}  
If all years are considered, there is a significant negative correlation between $A_i$ and $m_i$, with a Pearson's $r = -0.69$. However, if a change is assumed
at 2007, the negative correlation within the two populations is much smaller, $r=-0.37$ (1979-2006) and $r=-0.26$ (2007-2011). In neither case is the correlation
significantly different from zero. This suggests that the arctic climate has experienced a tipping point in 2007 to a new state where the mean ice area has shifted
to a lower level and the amplitude of the annual cycle has shifted to a higher level. The climatic transition is different from the recently proposed bifurcation from a mono-stable state to a bistable state[\cite{livina:2012,ditlevsen:2012}]. 

\section*{The Mean Seasonal Cycle}
\begin{figure}[H]
\begin{center}
\epsfxsize=8cm\epsffile{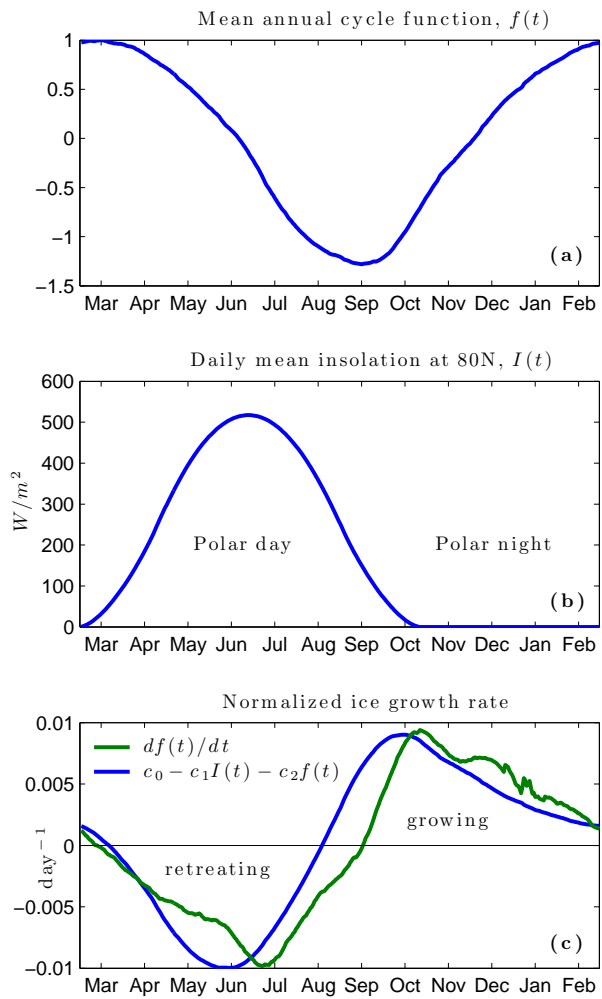}
\end{center}
\caption{The general annual cycle in Arctic sea ice area. (a) The cycle shown beginning at the time of maximum extent at the dawn of
the polar day. (b) shows the mean daily insolation at 80N. (c) The change in the normalized sea ice area (green) is a simple function of the
insolation and a melt rate proportional to the ice area itself.   
}
\label{f4}
\end{figure} 
The mean seasonal cycle function $f(t)$ of the ice area, is plotted again in figure 4(a) beginning the year with the dawn of the polar day. The change in this normalized ice area can 
only depend on the wind driven transport[\cite{ogi:2012,rigor:2002}], oceanic flow and SAT in an average sense, represented by a constant growth rate $c_0$. Radiative forcing is proportional to the insolation $I(t)$ over the arctic ocean and the amount of melting ice must be proportional to 
ice area $f(t)$ itself. Thus the simplest possible linear model is:  $df(t)/dt=c_0-c_1I(t)-c_2 f(t)$. The Arctic Oscillation[\cite{rigor:2004}] and other climate fluctuations on time scales longer than the correlation time of the noise $\epsilon_i(t)$ must be reflected in the annual variables $m_i$ and $A_i$.
Taking $I(t)$ to be the mean daily insolation at 80N (figure 4(b)), we see that the ice area is at its minimum on
September 8, right before beginning of the polar night, where it grows to its maximum at March 9, with the dawn of the polar day.
Figure 4(c) shows $df(t)/dt$ (in green) and $c_0-c_1I(t)-c_2 f(t)$ (in
blue) with $(c_0, c_1, c_2)=(0.0056, 0.00003, 0.01)\,  \mbox{day}^{-1}$. The agreement is surprisingly good. An important finding is that the accelerations
of the ice retreat in the polar day and the ice growth at polar night (slope of green curve in figure 4(c)) are very similar. This is consistent with flux calculations from reanalysis[\cite{perovich:2007}]  that the ice albedo 
feedback, which is only active during polar day is reflected in the the sea ice dynamics through the variables $A_i$ and $m_i$. 

\section*{Summary}
In summary, the Arctic sea ice shows a remarkably regular seasonal cycle where only the annual amplitude and mean change from year to year. This is remarkable for two
unrelated reasons: Firstly,
the variations in the total ice area covers regions with large variations and different trends[\cite{cavalieri:2012}]. Illustrative animations are provided by Google Earth[\cite{googleearth}]. Secondly, different changing factors influencing the ice growth and retreat have distinct seasonalities[\cite{serreze:2007,lu:2009}]. The decomposition
of the sea ice signal as done here, and the constancy of the normalized annual cycle, indicates that all these factors add up such that  the
amplitude is the strongest indicator of climate change. A significant tipping point was crossed in 2007 where the statistics permanently changed to a state of larger seasonal amplitude.   

\section*{Appendix: Methods}
\subsection*{Decomposing the sea ice area $x(t)$}
By averaging equation (1) over the year we estimate $m_i\approx \langle x_i(t)\rangle_t\pm \sigma/\sqrt{\tilde{n}}$, where $\sigma^2$ is the variance of the
noise $\epsilon(t)$ and $\tilde{n}=365/22$ is the effective number of independent points within a year. A posterior the relative intensity of the noise is calculated to $\sigma/m=0.02$,
thus the uncertainty is negligible. Likewise we obtain the amplitude from 
$\langle x_i(t)^2\rangle_t=m_i^2+A_i^2\langle f(t)^2\rangle_t +\sigma^2\Rightarrow A_i=\sqrt{\langle x_i(t)^2\rangle - m_i^2}$, where we have safely neglected the
$\sigma^2$ term, which gives a relative error of less than 0.004. The mean cycle function is obtained as $f(t)=\langle (x_i(t)-m_i)/A_i\rangle_i$, where $\langle . \rangle_i$ denotes
averaging over the 33 years. The uncertainty is of the order $\sigma/(\langle A_i\rangle_i\sqrt{33})=0.004$, thus also negligible. As a consistency check of equation (1), we
obtain $\langle f(t)\rangle_t/\sqrt{\langle f(t)^2\rangle_t}=0.0015\approx 0$. As to the effect of changing the beginning date for the year see the Supplementary material. 
\subsection*{On significance of change in amplitude}
Assuming $A_i$ to be an uncorrelated stochastic variable,
(the autocorrelation is -0.02 for the period 1979-2006), it is most likely that the arctic climate experienced a transition to a a new 
statistical state in 2007. A heuristic argument is based on the observation that the five highest values has been observed in the last five years of the record.
For any distribution for $A_i$, only using the observed independence, the probability of a coincidental occurrence of the last five consecutive years
having the five largest values out of 33 years is 1/(number of ways to pick 5 years of 33 years)=$1/{33 \choose 5}= 4.2\times 10^{-6}$, thus very unlikely. 
Note that this argument for a changing statistics is not rigorously correct, since any sequence of occurrences of, say, the five
largest values in 33 drawings is equally unlikely without indicating change in statistics. The fact that the criterion of change is derived from the data and not
independent flaws the argument. Note also that the argument cannot be applied to the statistics of the mean $m_i$ where
the value for 2006 is smaller than values for both 2008 and 2009.
\subsection*{Change point analysis}
The change point analysis[\cite{barry:1993}] is based on Bayesian inference, where an unknown number of change points in the data series are assumed. In between 
the change points the variables are drawn from a gaussian distribution with constant mean and variance. The probability of a change point can be calculated recursively
from the conditional probability of the next data point given the maximum likelihood parameters determining the distribution of the previous data. The prior distribution
for a change point is taken to be uniform. The numerical algorithm is taken from ref.[\cite{adams:2007}].


\end{document}